\documentclass[apj]{emulateapj}
\usepackage{graphicx}
\usepackage{amsmath}
\usepackage{natbib}

\newcommand{\om}{\Omega_m}
\newcommand{\ok}{\Omega_k}
\newcommand{\ox}{\Omega_x}
\newcommand{\oll}{\Omega_\Lambda}
\newcommand{\orc}{\Omega_{r_c}}
\newcommand{\obh}{\Omega_{b}h^2}
\newcommand{\lsim}{\mbox{$\:\stackrel{<}{_{\sim}}\:$} }

\newcommand{\ols}{\bar{\Omega}_{ls}^\phi}

\shorttitle{Scrutinizing the viability of candidate dark energy scenarios}
\shortauthors{Davis, M\"ortsell, Sollerman et al.}

\begin{document}

\title{Scrutinizing exotic cosmological models\\ using ESSENCE supernova data combined with other cosmological probes}

\author{
T.~M. Davis\altaffilmark{1}, 
E.~M\"ortsell\altaffilmark{2}, 
J.~Sollerman\altaffilmark{1,2}, 
{A.~C.~Becker}\altaffilmark{3},  
{S.~Blondin}\altaffilmark{4}, 
{P.~Challis}\altaffilmark{4},  
{A.~Clocchiatti}\altaffilmark{5},  
{A.~V.~Filippenko}\altaffilmark{6},  
{R.~J.~Foley}\altaffilmark{6},  
{P.~M.~Garnavich}\altaffilmark{7}, 
{S.~Jha}\altaffilmark{8},  
{K.~Krisciunas}\altaffilmark{7,9}, 
{R.~P.~Kirshner}\altaffilmark{4}, 
{B.~Leibundgut}\altaffilmark{10}, 
{W.~Li}\altaffilmark{6}, 
{T.~Matheson}\altaffilmark{11}, 
{G.~Miknaitis}\altaffilmark{12},  
{G.~Pignata}\altaffilmark{5},  
{A.~Rest}\altaffilmark{13}, 
{A.~G.~Riess}\altaffilmark{14,15},  
{B.~P.~Schmidt}\altaffilmark{16}, 
{R.~C.~Smith}\altaffilmark{13},  
{J.~Spyromilio}\altaffilmark{10}, 
{C.~W.~Stubbs}\altaffilmark{4,17}, 
{N.~B.~Suntzeff}\altaffilmark{9,13}, 
{J.~L.~Tonry}\altaffilmark{18}, 
{W.~M.~Wood-Vasey}\altaffilmark{4}, 
{A.~Zenteno}\altaffilmark{19}
}
\email{tamarad@dark-cosmology.dk}

\altaffiltext{1}{Dark Cosmology Centre, Niels Bohr Institute, University of Copenhagen, Juliane Maries Vej 30, DK--2100 Copenhagen~\O, Denmark; tamarad@dark-cosmology.dk; jesper@dark-cosmology.dk}
\altaffiltext{2}{Department of Astronomy, Stockholm University, AlbaNova, S-106 91 Stockholm, Sweden;  edvard@astro.su.se}
\altaffiltext{3}{Department of Astronomy, University of Washington, Box 351580, Seattle, WA 98195-1580} 
\altaffiltext{4}{Harvard-Smithsonian Center for Astrophysics, 60 Garden Street, Cambridge, MA 02138} 
\altaffiltext{5}{Pontificia Universidad Cat\'olica de Chile, Departamento de Astronom\'ia y Astrof\'isica, Casilla 306, Santiago 22, Chile} 
\altaffiltext{6}{Department of Astronomy, 601 Campbell Hall, University of California, Berkeley, CA 94720-3411} 
\altaffiltext{7}{Department of Physics, University of Notre Dame, 225 Nieuwland Science Hall, Notre Dame, IN 46556-5670} 
\altaffiltext{8}{Kavli Institute for Particle Astrophysics and Cosmology, Stanford Linear Accelerator Center, 2575 Sand Hill Road, MS 29, Menlo Park, CA 94025} 
\altaffiltext{9}{Department of Physics, Texas A\&M University, College Station, TX 77843-4242} 
\altaffiltext{10}{European Southern Observatory, Karl-Schwarzschild-Strasse 2, D-85748 Garching, Germany} 
\altaffiltext{11}{National Optical Astronomy Observatory, 950 North Cherry Avenue, Tucson, AZ 85719-4933} 
\altaffiltext{12}{Fermilab, P.O. Box 500, Batavia, IL 60510-0500} 
\altaffiltext{13}{Cerro Tololo Inter-American Observatory, National Optical Astronomy Observatory, Casilla 603, La Serena, Chile} 
\altaffiltext{14}{Space Telescope Science Institute, 3700 San Martin Drive, Baltimore, MD 21218} 
\altaffiltext{15}{Johns Hopkins University, 3400 North Charles Street, Baltimore, MD 21218} 
\altaffiltext{16}{The Research School of Astronomy and Astrophysics, The Australian National University, Mount Stromlo and Siding Spring Observatories, via Cotter Road, Weston Creek, PO 2611, Australia} 
\altaffiltext{17}{Department of Physics, Harvard University, 17 Oxford Street, Cambridge, MA 02138} 
\altaffiltext{18}{Institute for Astronomy, University of Hawaii, 2680 Woodlawn Drive, Honolulu, HI 96822} 
\altaffiltext{19}{Department of Astronomy, University of Illinois at Urbana-Champaign, 1002 West Green St, Urbana, IL 61801-3080}

\begin{abstract}
The first cosmological results from the ESSENCE supernova survey
\citep{WV07} are extended to a wider range of cosmological models
including dynamical dark energy and non-standard cosmological models.
We fold in a greater number of external data sets such as the recent
Higher-$z$ release of high-redshift supernovae \citep{riess07} as well
as several complementary cosmological probes. Model
comparison statistics such as the Bayesian and Akaike information
criteria are applied to gauge the worth of models. These statistics favor 
models that give a good fit with fewer parameters.  

Based on this analysis, the preferred cosmological model is the
flat cosmological constant model, where the expansion history of the
universe can be adequately described with only one free parameter
describing the energy content of the universe.  Among the more exotic models that provide good fits to the data, we note a preference for models whose best-fit parameters reduce them to the cosmological constant model.
\end{abstract}


\keywords{cosmology: observations ---
supernovae : general}


\section{Introduction}\label{intro}
Ever since Type Ia supernova (SN Ia) measurements first indicated an accelerating expansion of  the universe \citep{riess98,perlmutter99}, the focus of cosmology has shifted dramatically.  Over the last several years, the primary aim of many cosmological observations has been to discover the reason for this accelerated expansion.  The name we often give to the unknown cause of the acceleration is ``dark energy.''  In this context, dark energy represents not only the possibility of a hitherto undiscovered component of the energy density of the universe but also the possibility that the standard models of gravity and/or particle physics require revision. 

The ESSENCE (``Equation of State: SupErNovae trace Cosmic Expansion'') supernova survey is an ongoing project that aims to measure the equation-of-state parameter of dark energy to better than 10\% \citep{krisciunas05,sollerman06,miknaitis07}.
In this paper, we use the first cosmological results paper from the ESSENCE supernova survey \citep[][hereafter WV07]{WV07}, where distances and redshifts for a large sample of newly discovered high-redshift SNe Ia are reported. Moreover, WV07 performed consistent light-curve fitting of not only the ESSENCE sample of supernovae but also the local sample \citep{hamuy96,riess99,jha06} as well as the SN Ia data released by the SuperNova Legacy Survey \citep[SNLS;][]{astier06}.  Combining this with constraints from baryon acoustic oscillations \citep[BAOs;][]{eisenstein05}, WV07 placed constraints on the dark energy equation-of-state parameter of $w=-1.07\pm0.09$ (statistical $1\sigma$)$\pm0.13$ (systematic) and on the matter density of $\om=0.267^{+0.028}_{-0.018}$ (statistical 1$\sigma$) for a flat universe.\\

To extend the analysis presented by WV07 we have

(1) added the 30 SNe Ia detected at $0.216\le z\le1.755$ by the {\it Hubble Space Telescope (HST)} as reported by \citet{riess07};

(2) included further constraints from a wider range of complementary observations;

(3) allowed for more complex cosmological models by both relaxing the assumption of  a flat universe and testing non-standard models inspired by new fundamental physics; and

(4) applied model-comparison statistics 
to decide on the model best preferred by the current data.  

Today's SN Ia results are accommodated in what has become the concordance cosmology \citep{spergel03}, flanked by constraints on the matter density, $\Omega_{m}$, from large-scale structure measurements and on the flatness of space from cosmic microwave background (CMB) measurements \citep{spergel06}.
The concordance cosmology is dominated by dark energy, $\Omega_{x}\approx 0.7$, with present evidence being consistent with 
Einstein's cosmological  
constant, $\Lambda$ \citep{einstein17,zeldovich67,padmanabhan03}. 
We refer to the model with a cosmological constant as the standard model, or $\Lambda$ model.

However, uncertainty remains over whether the cosmological constant is indeed the complete description of dark energy or whether the dark energy might have more complex behavior.    This question is motivated by the enormous discrepancy between the theoretical prediction for 
the cosmological constant and its measured value \citep{weinberg89}.  That we apparently live in an era when $\Omega_{\Lambda}$ and  $\Omega_{m}$ are almost equal, known as the ``coincidence problem,'' also suggests that we may have an incomplete cosmological model.
Thus, a variety of suggestions for new physics have emerged.  

Some of these suggestions take the form of variations of the equations of general relativity \citep[e.g., the Dvali-Gabadadze-Porrati (DGP) model;][]{dgp00}, while others invoke more complex and evolving forms of dark energy \citep[e.g., quintessence;][]{caldwell98}. 
A basic way 
to explore more complex models is to parameterize the dark energy by an equation-of-state parameter $w$, relating the dark-energy pressure, $p$, to its density, $\rho$, via $p\,=\,w\,\rho\, c^{2}$. (Hereafter we set $c = 1$.) 
This parameter may be time variable and characterizes how the energy density evolves with the scale factor, $a$: 
$\rho~\propto~a^{-3(1+w)}$. 

This paper is organized as follows.  In Sect.~\ref{sect:modelcomparison} we discuss information criteria in the context of cosmological model selection. Section~\ref{sect:data} details the data sets used, their individual systematics and assumptions, and our method of combining them.  Section~\ref{sect:models} describes each model in turn and assesses which is preferred by the data, the results of which are discussed in Sect.~\ref{sect:results}.  

\section{Model selection vs fitting parameters}\label 
{sect:modelcomparison}

Parameter fitting and goodness-of-fit (GoF)\footnote{Defined as GoF = $\Gamma(\nu/2,\chi^2/2)/\Gamma(\nu/2)$, where $\Gamma(\nu/2,\chi^2/2)$ is the incomplete gamma function and $\nu$ is the number of degrees of freedom.    It gives the probability of obtaining data that are a worse fit to the model, assuming that the model is correct.} 
 tests alone are not effective ways to decide between possible models.  These statistical measures are based on the assumption that the underlying model is the correct one.  The $\chi^2$ statistic can test the validity of a particular model, but comparing relative likelihoods based on the $\chi^2$ values of different models does not properly account for the structural differences between them.

In other words, $\chi^2$ statistics are good at finding the best parameters in a model but are 
insufficient for deciding whether the model itself is the best one.  
One might be tempted to prefer the model that gives the best fit to the data, defined as the lowest $\chi^2$.  However, this does not account for the relative complexity of the models.  To give a blatant example, a 10th-order polynomial will {\em always} give an equal or better fit than a straight line to any data set, but this does not mean that any of the extra eight coefficients have any  significance.  It just means that a model with more parameters will generally give an improved fit (always, if the simpler model is a subclass of the more complex one).  

Moreover, even though many cosmological models can be expressed in terms of a ``$w$'' that describes the dynamical behavior of dark energy, the different functional parameterizations used by different models mean that they are not referring to the same thing.    Consequently, the value of $w$ that the data prefer is integrally related to the model used in a fit \citep[e.g.][]{zhao06}.  This difficulty makes it unwise to compare different models by simply considering likelihood contours or best-fit parameters.
For example,  the constant, $w$, that appears in the standard dark-energy model (Sect.~\ref{sect:w}) is not the same parameter as the constant, $w_0$, that appears in the $w(a)$ parameterization of the variable dark-energy model (Sect.~\ref{sect:wa}).  So if the best-fit value of $w_0$ drifts away from $-1$ it does not rule out $w=-1$.

Instead we turn to information criteria (IC) to assess the strength of
models.  These statistics favor models that give a good fit with fewer parameters.
In this paper we use the Bayesian information criterion \citep[BIC;][]{schwarz78}
and the Akaike information criterion \citep[AIC;][]{akaike74} to select the best-fit
models.
\citet{liddle04}  examines the use of information criteria in the context of cosmological observations,  and we follow his prescription here.  
Previous explorations into AIC and BIC in a cosmological context include \citet{godlowski05}, \citet{sydlowski06a}, \citet{szydlowski06b}, \citet{magueijo06}, \citet{mukherjee06} and \citet{biesiada07}.

The BIC \citep[also known as the Schwarz information criterion;][]{schwarz78} is given by
\begin{equation} {\rm BIC} = -2 \ln{\cal L} + k \ln N, \end{equation}
where ${\cal L}$ is the maximum likelihood,
$k$ is the number of parameters, and $N$ is the number of data points used in the fit.  Note that for Gaussian errors, $\chi^2 = -2 \ln{\cal L}$, and the difference in BIC can be simplified to $\Delta BIC=\Delta\chi^2+\Delta k \ln N$.  A difference in BIC ($\Delta$BIC) of 2 is considered positive evidence against the model with the higher BIC, while a $\Delta$BIC of 6 is considered strong evidence \citep{liddle04}.

The AIC \citep{akaike74},
\begin{equation} {\rm AIC} = -2 \ln {\cal L} + 2 k ,\end{equation}
gives results similar to the BIC approach, although the AIC is more lenient on models with extra parameters for any reasonably sized data set ($\ln N > 2$).  As mentioned in \citet{liddle07}, \citet{sugiura78} give a version of the AIC corrected for small sample sizes, ${\rm AIC}_c = {\rm AIC} + 2k(k-1)/(N-k-1)$, which is important when N/k \lsim 40 \citep{burnham02,burnham04}.   The correction is negligible in our case ($\sim0.06$).

Both tests can be applied to unrelated models (the simpler model need not be nested in the more complex model).   
These criteria make an attempt to quantify Occam's razor.  When two models fit the data equally well, the simpler model (the one with fewer free parameters) is preferred.  

A poor information criterion result can arise in two ways: 
(1) when the model is a poor fit to the data, regardless of the number of free parameters; in this case a large reduced $\chi^2$ value indicates that the model does not explain the data; and  
(2) when the data are too poor to constrain the extra parameters in the model; such a model would be disfavored by information criteria if a simpler model is available, but it may well be that with improved data the more complex model becomes preferred.\footnote{For example, general relativity would fail to rival Newtonian gravity if the only experiment available were 
dropping balls from the leaning tower of Pisa, 
but general relativity would become necessary when data on the precession of Mercury's orbit became available.}  
Thus, information criteria alone can at most say that a more complex model is {\em not necessary} to explain the data.  In this paper, we find this to be the current situation for dynamical dark-energy models.  

The simple prescription for information criteria above is limited.  A more in-depth analysis of the improvement gained by more complex models would not simply count parameters, but would consider how much the allowed volume in data space increases due to the addition of extra parameters (i.e. how much more flexible the model becomes), as well as any correlations between the parameters.   
Bayesian model selection is a technique that takes this into account using Bayesian evidence, i.e., the average likelihood of a model over its prior parameter ranges.   \citet{saini04} pioneered the use of Bayesian evidence in cosmology and \citet{liddle06} analyze a variety of different parameterizations of dark energy, $w$, finding that the standard cosmological constant 
 remains the favored model.  Other studies include \citet{elgaroy06}.
For this paper, we have used the first approximation provided by the IC without calculating the full Bayesian evidence.  This simpler version is sufficient for our purposes, and when systematic errors dominate the uncertainty in the data (as has now become true for SN data sets), further statistical analysis becomes unwarranted. 
IC also require no assumptions for the prior or the metric on the space of model parameters.

\section{Data}\label{sect:data}

In order to test the different models, we have used observational data from a variety of sources described below.

\subsection{Type Ia Supernovae (SNe Ia)}

As mentioned in the introduction, one of the prime new data sets in this work is the SN Ia data from the ESSENCE project. The ESSENCE project is a ground-based survey designed to detect and monitor about 200 SNe~Ia in the redshift range $z = 0.2-0.8$.

The strategy and implementation of this project are described by \citet{miknaitis07}, and the goal of the completed survey is to constrain $w$ with a precision of $\sim10\% $. With the recent release of the first 4 years of data and their cosmological implications (WV07), we are now in an excellent position to probe further cosmological models. 

Another large, high-quality supernova search, the SNLS, recently published an extensive and homogeneous first-year data set \citep{astier06}.
In combining the two data sets we have leaned on the light-curve fitting performed by WV07, who fit all SNe from these different data sets with the same light-curve fitter, MCLS2k2 \citep{jha07}. 
Following WV07, the uncertainty added due to the intrinsic diversity of SNe Ia is 0.10 mag (assuming a peculiar velocity uncertainty of 400 km~s$^{-1}$). 
We have used the data from Table 9 in WV07, including only those SNe for which the light-curve fits passed the quality cuts.  That includes 60 ESSENCE supernovae, 57 SNLS supernovae, and 45 nearby supernovae.

We also incorporate the new data release of 30 SNe~Ia, at $0.216\le z\le 1.755$ detected by {\it HST} and classified as ``gold'' supernovae by \citet{riess07}.
Such high-$z$ data are particularly useful for this analysis because the SN Ia constraints on the {\em evolution} of $w$ are improved as the range of redshifts is extended. 
We adopted the local supernovae that these samples had in common in order to normalize the luminosity distances of the samples, and we included the uncertainty in this normalization in the distance errors for the {\it HST} SNe~Ia.  

Ideally these two data sets should both be generated using the same light-curve fitter, in which case no normalization would be required.  This is in progress, and in the interim we provide the combined data set as used in this paper.\footnote{The data are available at \url{http://www.dark-cosmology.dk/archive/SN}, \url{http://braeburn.pha.jhu.edu/$\sim$ariess/R06} and \url{http://www.ctio.noao.edu/essence}.  When used, please cite WV07, \citet{riess07} and \citet{astier06} in addition to this paper. These Web sites will be updated when the self-consistent light-curve-fitting is complete.}

We note that any statistical analysis of the type we perform here may be limited by systematic errors in the data. 
 Possible sources of systematic error in the supernova data include local velocity structures \citep{jha07,zehavi98} 
and the treatment of dust \citep{WV07,astier06}. 
 WV07 concentrate much of their discussion on the analysis of systematic errors and how to minimize them.  They calculate that the systematic error in supernova data gives a maximum uncertainty in $w$ of 0.13 for the flat, constant-$w$ model when combined with BAO constraints.

\subsection{Cosmic Microwave Background (CMB)}
The characteristic angular scale of the first peak in the CMB
anisotropy spectrum is given by,
\begin{equation}
\theta_A \equiv \frac{r_s(z_{ls})}{D_A(z_{ls})}\ ,
\end{equation}
where $r_s(z_{ls})$ is the comoving size of the sound horizon at last
scattering [roughly proportional to $(\om H_0^2)^{-1/2}$] and
$D_A(z_{ls})$ is the comoving angular distance to the last-scattering
surface.

Following the prescription given by \citet{doran02} and \citet{page03},
we have converted the {\it WMAP} three-year result
\citep{spergel06} on the location of the first peak to 
a reduced distance to the last-scattering surface,
\begin{equation} R=\sqrt{\frac{\om}{|\ok|}}\; S_k\left[H_0\sqrt{|\ok|}\int_0^{z_{\rm ls}}\frac{dz}{H(z)}\right]=1.71\pm 0.03,\end{equation}
where $S_k(x)=\sin{x}, x$, and $\sinh{x}$ for $k=+1, 0$, and $-1$, respectively.\footnote{Other calculations of the shift parameter include \citet{wang06}, who find $R=1.70\pm0.03$, and this is used by, for example, \citet{alam06} and \citet{liddle06}.  \citet{elgaroy07} find $R=1.71\pm0.03$.  We calculated the value independently.  In a new paper \citet{wang07} find $R=1.71\pm0.03$ when allowing nonzero cosmic curvature.  The small difference has a negligible effect on the results.} 

The value of this parameter is somewhat model dependent.  For example, it changes slightly when massive neutrinos are included \citep{elgaroy07}, and it is weakly dependent on the density of the dark energy at the last scattering surface \citep[through the $\ols$ term: see][]{doran02}.  Using $R=1.71$ therefore artificially excludes some models with, for example, strongly varying dark energy.  However, constraints from big bang nucleosynthesis (BBN) rule out models with vastly different expansion dynamics from the standard model at the time of nucleosynthesis \citep{carroll02, steigman06, wright07}, so this is not an unreasonable exclusion.  The BBN results may change if not only gravitational dynamics but also particle physics processes changed in any of the models.

The robustness of the shift parameter is tested in \citet{elgaroy07} compared to fitting the full CMB power spectrum, and they confirm that it is an accurate measure for non-standard cosmologies such as those tested here.  Degeneracies that arise from using $R$ rather than fitting the full CMB power spectrum are well constrained by other data such as BAOs and SNe. 

\subsection{Baryon Acoustic Oscillations (BAO)}
Similar to the use of the angular scale of the first peak in the CMB
spectrum, we can use the measurement of the peak at $\sim 100
h^{-1}$ Mpc of the large-scale correlation function of luminous red
galaxies in the Sloan Digital Sky Survey
\citep{eisenstein05} to further constrain cosmological parameters.

The large-scale correlation function is a combination of the
correlations measured in the radial (redshift space) and the
transverse (angular space) direction, and thus
the relevant distance measure is the so-called dilation scale,
\begin{equation}
  D_V(z)=[D_A(z)^2 z/H(z)]^{1/3}\ ,
\end{equation}
at the typical redshift of the galaxy sample, $z=0.35$.
The absolute scale of the BAO is given
by the sound horizon at last scattering, and the dimensionless combination $A(z)=D_V(z)\sqrt{\om
H_0^2}/z$ is well
constrained by the BAO data to be 
\begin{equation} A(0.35)=D_V(0.35)\frac{\sqrt{\om H_0^2}}{0.35}=0.469\pm 0.017.\end{equation}

\subsection{Not Used}

There are additional sources of observations, which are potentially important for constraining cosmological parameters, that we have not included in this analysis. We have, for example, chosen to omit any information from
distant gamma-ray bursts \citep{ghirlanda04} since such data are rather controversial \citep{mortsell05,friedman05}. We have also not used the constraints from X-ray data on relaxed galaxy clusters \citep{allen04}, although this method may well become important in the future. The weak lensing data from the Canada-France-Hawaii Telescope \citep[CFHT;][]{hoekstra06} are an additional source of data that we have omitted here.  At least some of these additional sources of information will become increasingly important as the data improve. 

Apart from geometrical probes, as a consistency check we have used
large-scale structure constraints on the growth factor as measured by the Two Degree Field Galaxy Redshift Survey. The
constraints on the models from the growth factor are similar, but
weak, compared to the BAO constraints. Since the computation of the growth
factor in modified gravity models is potentially very complicated, we
have chosen not to include those constraints in the final analysis.

\subsection{Combining the Constraints}
Both the distance to the last-scattering surface derived from the CMB and the BAO constraints depend on the baryon density ($\obh$) and its uncertainty, and for this we use the value obtained from {\it WMAP}. However, the dependence of BAO on baryon density is weak [$\propto (\obh)^{-0.08}$ \citep{eisenstein05}], and for distance-related parameters, BAOs and CMB are independent.
Since the SNe~Ia, CMB, and BAOs are effectively independent
measurements, we can combine our results by simply multiplying the
likelihood functions.


\begin{deluxetable}{llll}
\tablewidth{0pc}
\tablecolumns{4}
\tablecaption{Summary of models}
\tablehead{
\colhead{Model}  &
\colhead{Abbrev.\tablenotemark{a}} &
\colhead{Parameters\tablenotemark{b}} &
\colhead{Section}
}
\startdata
  Flat cosmo. const. & F$\Lambda$ & $\om$   & \ref{sect:FL}  \\ 
 Cosmological const. &  $\Lambda$ & $\om$, $\oll$   & \ref{sect:L}   \\ 
 Flat constant $w$ &         Fw & $\om$, $w$   & \ref{sect:Fw}  \\ 
      Constant $w$ &          w & $\om$, $\ok$, $w$     & \ref{sect:w}  \\ 
         Flat w(a) &        Fwa & $\om$, $w_0$, $w_a$    & \ref{sect:wa}  \\ 
               DGP &        DGP & $\ok$, $\orc$   & \ref{sect:dgp}   \\ 
          Flat DGP &       FDGP & $\orc$    & \ref{sect:fdgp}  \\ 
       Cardassian &         Ca & $\om$, $q$, $n$    & \ref{sect:ca}  \\ 
Flat Gen. Chaplygin &       FGCh & $A$, $\alpha$  & \ref{sect:gch}   \\ 
    Gen. Chaplygin &        GCh & $\ok$, $A$, $\alpha$    & \ref{sect:gch} \\ 
    Flat Chaplygin &        FCh & $A$   & \ref{sect:ch}   \\ 
         Chaplygin &         Ch & $\ok$, $A$     & \ref{sect:ch}  
\enddata
\tablenotetext{a}{The abbreviations used in Fig.~\ref{fig:daic}.}
\tablenotetext{b}{The free parameters in each model.  Note that when fitting the SN Ia data we also fit an additional parameter, $\cal{M}$, for the normalization of SN magnitudes.  We include this in the number of degrees of freedom and in $k$, but have not listed it here as a parameter in each model.}
\label{t:params}
\end{deluxetable}
\begin{figure}
 \plotone{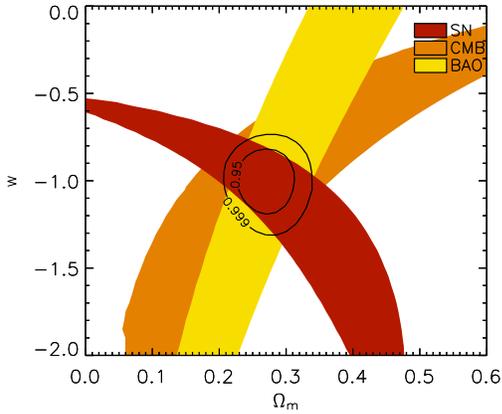}
   \caption{Flat dark-energy model: a flat universe with constant $w$ (Sect.~\ref{sect:flatde}).  The constraint from each of the observational probes is shown by shaded contours (according to the legend).  These are all 95\% confidence intervals for two parameters.  The combined contours (95\% and 99.9\% confidence intervals) are overlayed in black.  The complementarity of the different observational probes is clearly demonstrated in the differing angles of the overlapping contours.  
The combined data form a clear preference around the cosmological constant model ($w=-1$).  Despite the extra freedom afforded by allowing the dark energy to have an equation-of-state parameter that differs from $-1$, the data do not show any indication that this freedom is required.} 
\label{fig:smfcomb} 
\end{figure}

\section{Models}\label{sect:models}

Hitherto, all measurements of $w$ have been consistent with a cosmological  constant, $w=-1$ \citep[e.g.,][]{garnavich98,tonry03,knop03,hannestad04, eisenstein05,astier06,spergel06,riess07,WV07}.  
 However, as discussed in the introduction, this is not an  unproblematic conclusion. 
Given the theoretical difficulties in predicting a cosmological constant with the right vacuum energy density, 
 a variety  of suggestions for other new physics have emerged.  
Many models use  evolving scalar fields: so-called quintessence models, 
which allow a time-varying equation of state to track  the matter density. In such models, the time-averaged absolute value  of $w$ is likely to differ from unity.  Many of these have been tested by other authors; for example, \citet{wilson06} show that one specific version of quintessence, the inverse power-law potential \citep{peebles88}, has a best-fit solution consistent with the cosmological constant model.  
Many other models including  all kinds of exotica have been proposed, such as $k$-essence, domain walls,  frustrated topological defects and extra dimensions. 
\citet{padmanabhan03} gives a review of dark energy and its alternatives.  

Models can be broadly classed into two groups: (1) those that invoke some form of extra component to the composition of the universe (such as dark energy or quintessence), or (2) those that invoke a variation in the equations governing gravity.  In some cases the two are interchangeable descriptions of a single theory. 

Here we choose several of the most popular models discussed in the literature and examine whether they are consistent with the data currently available to us:

1. Standard dark-energy models, including varying $w$.

2.~Dvali-Gabadadze-Porrati (DGP) brane world model. 

3. Cardassian expansion.

4. Chaplygin gas.

In the following sections we outline the basic equations governing the evolution of the expansion of the universe in each of the different models, calculate the best-fit values of their parameters, and find their  $\Delta$AIC and $\Delta$BIC values.  The models used and the parameters that describe each model are summarized in Table~\ref{t:params}.  For some relevant models we plot the likelihood contours of their parameters (see Figs.~\ref{fig:smfcomb}--\ref{fig:schcomb}).  We show how the magnitude-redshift evolution of each model compares to the supernova data in Figure~\ref{fig:hd} and how their best fits match the value of $R$ measured from the CMB data and the value of $A$ measured from the BAO data in Figure~\ref{fig:cmb_bao}.   The information criteria results are summarized in Table~\ref{t:aic} and shown in Figure~\ref{fig:daic}.  

All the contours in Figures~\ref{fig:smfcomb} -- \ref{fig:schcomb} represent two-parameter confidence intervals and the uncertainties quoted in the text are the 95\% confidence level for one parameter.

\subsection{Dark-Energy Models with {\bf Constant} Equation of State}

The dark-energy models with a constant equation-of-state parameter, $w$, are described by the following equation relating Hubble's constant to the scale factor, $a$:

\begin{equation}{H^2 \over H_0^2}={\om\over a^3}+{\ok\over a^2}+{\ox\over a^{3(1+w)}}\, ,\label{eq:darkenergy}\end{equation}
where $\om$ is the current value of the normalised matter density, the curvature of the universe is given by $\ok=1-\ox - \om$, and $\ox$ is the current value of the normalised dark-energy density. 

\subsubsection{Flat, Cosmological Constant Model (Flat $\Lambda$)}\label{sect:FL}

The standard cosmological model is the $\Lambda$CDM model, which  
invokes $w=-1$ at all times, in the form of a  
cosmological constant, $\ox=\oll$.  The simplest version of this  
model assumes a flat universe ($\oll=1 - \om$), so
\begin{equation}{H^2 \over H_0^2}={\om\over a^3}+(1-\om) \, ,\label{eq:Lcdm}\end{equation}
which only depends on one parameter.  
Our best-fit value is
$$\om=0.27 \pm 0.04.$$ 

This has the lowest AIC and BIC of all models tested, so $\Delta$AIC and $\Delta$BIC are measured with respect to this model (Table~\ref{t:aic}).

\subsubsection{The Cosmological Constant Model ($\Lambda$)}\label{sect:cc}\label{sect:L}

Allowing for deviations from flatness allows one extra degree of  
freedom, 
\begin{equation}{H^2 \over H_0^2}={\om\over a^3}+{\ok\over a^2}+{\oll}\, ,\end{equation}
(recalling $\ok=1-\oll - \om$).

\subsubsection{The Flat Dark-Energy Model (Flat Constant $w$)}\label{sect:flatde}\label{sect:Fw}

Allowing instead for different values of $w$, but maintaining the  
constraint of flatness, we have
\begin{equation} {H^2 \over H_0^2}={\om\over a^3}+{\ox\over a^{3(1+w)}}\, ,\end{equation}
where $\ox = 1-\om$.  We plot the likelihood contours in Figure~\ref{fig:smfcomb}.  
The best-fit parameters are,
$$\om = 0.27\pm0.04,\quad w = -1.01\pm0.15 \, .$$ 

\subsubsection{The Standard Dark-Energy Model (Constant $w$)}\label{sect:standardde}\label{sect:w}

Relaxing the constraint of flatness, we fit the most general form of  
constant-equation-of-state dark energy using all three independent  
parameters of Eq.~\ref{eq:darkenergy}.  

The AIC and BIC for these four models suggest that the simplest model  
adequately explains the data, and there is little evidence supporting  
the inclusion of extra parameters.

\begin{figure}
   \plotone{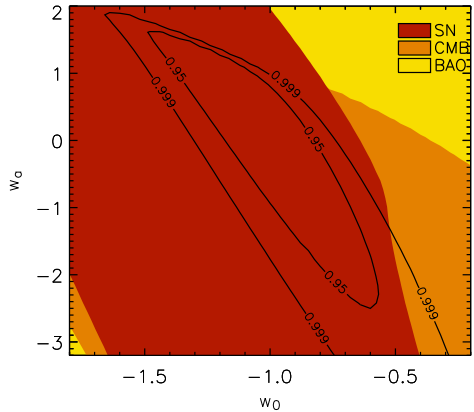}
   \caption{Flat, variable $w(a)$ model (Sect.~\ref{sect:wa}).  The contours are the same as in Fig.~\ref{fig:smfcomb}.  The parameters of this model are very poorly constrained by the current data.} 
   \label{fig:smfullcomb} 
\end{figure}

\subsection{Dark-Energy Models with {\bf Variable} Equation of State}\label{sect:variablede}

Allowing the dark-energy equation-of-state parameter to vary as the universe  
evolves adds additional degrees of freedom to the model.  The Friedmann equation for a varying $w$ is given by Eq.~\ref{eq:darkenergy} with the following replacement:
\begin{equation} a^{3(1+w)} \rightarrow \exp \left(3\int_a^1\frac{1+w(a^\prime)}{a^\prime}da^\prime\right).\label{eq:aarrow} \end{equation}
Below we discuss one possible parameterization of $w$, which is linear in scale factor.  However, it is important to note that the time variation of $w$ can be parameterized in many ways \citep[e.g.,][]{hannestad04,barnes05,calvo06,huterer06, riess07}, and choosing a particular parameterization enforces a particular form of time evolution on the model, which may not be appropriate.  Giving an analytic form to the time evolution of $w$ can act like a strong prior and may give misleading results \citep[for further discussion see][]{bassett04,riess07}.  In what follows, we consider what has become the most-common parameterization of $w$ but caution that a non-parametric approach could be preferred.  Promising non-parametric techniques include a type of principal component analysis that allows the reconstruction of cosmological features such as $\rho(z)$, $w(z)$, and their derivatives, as a function of redshift  \citep[e.g.,][]{linderhuterer05,huterer05,huterer06,riess07}.

\subsubsection{Standard Parameterization (Flat $w_a$)}\label{sect:wa}
Using the parameterization $w(a)=w_0+w_a(1-a)$ \citep{chevallier01,linder03}, Eq.~\ref{eq:aarrow} simplifies to
\begin{equation} a^{3(1+w_0)} \rightarrow a^{3(1+w_0+w_a)}e^{3w_a(1-a)}. \end{equation}
When we fit for this model we assume a flat universe, although loosening this constraint does not change the results considerably. 
The best-fit parameters are
$$\om= 0.27 \pm 0.04, \quad w_0=-1.1^{+0.4}_{-0.3}, \quad w_a=0.8^{+0.8}_{-2.4}. $$

The large flexibility of this model means that it is poorly constrained by current data (see Fig.~\ref{fig:smfullcomb}).

This is also the model that is used in the Dark Energy Task Force  (DETF) figure of merit \citep[FoM;][]{DETFastroph06}.  The FoM is given by the inverse of the area of the 95\% confidence interval in the $w_0$-$w_a$ plane.  The smaller the area (thus the larger the FoM), the better the discriminating ability of the experiment.  We calculate that the data sets used here have an FoM of $\sim 1$.

\begin{figure}
   \plotone{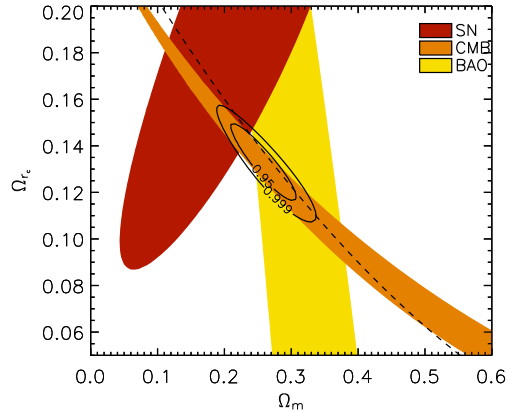}
   \caption{General DGP model (Sect.~\ref{sect:generaldgp}).  The dashed line shows the flat model.  Here the contours from the different observational constraints disagree and the model is thus strongly disfavored. } \label{dgpcomb}   
   \label{fig:dgp} 
\end{figure}


\subsection{The DGP Models}

The DGP models \citep{dgp00} arise from a  
class of brane-related theories in which gravity leaks out into the  
bulk at large distances, resulting in the possibility of accelerated  
expansion.

Notably, this theory provides for an accelerating universe without  
adding any extra parameters, two parameters being sufficient to  
define the model. 
\citet{lue06} shows how the growth of large-scale structure proceeds in the DGP model, but the position of the BAO peak is not expected to be influenced by this modification \citep{yamamoto06}. 

\subsubsection{DGP Model}\label{sect:generaldgp}\label{sect:dgp}
The general DGP model is governed by the equation
\begin{equation} {H^2 \over H_0^2}={\ok\over a^2}+
\left(\sqrt{{\om\over a^3}+\orc}+\sqrt{\orc}\right)^2 \, ,\end{equation}
where $\om = 1-\ok -2\sqrt{\orc}\sqrt{1-\ok}$.  The parameter $r_c$ is the length scale beyond which gravity leaks out into the bulk, and $\orc$ is related to this length scale by $\orc=1/(4r_c^2H_0^2)$.

As is evident from Figure~\ref{fig:dgp}, the overlapping region that is preferred by the supernova and BAO  data seems to be inconsistent with the CMB data.   
This concurs with the analysis of the DGP model by \citet{fairbairn06}. However, it should be noted that the model cannot be ruled out based on this observation alone since the GoF is $\sim 0.2$; that is, assuming that the underlying model is indeed DGP, the probability of finding an even worse fit is $\sim 20\,\%$.  

\vspace{5mm}
\subsubsection{Flat DGP Model}\label{sect:fdgp}
The flat DGP model can be considered a more constrained version of the general DGP model.  It has only one parameter to fit, $\orc$, and serves as an illustrative example of the power of information
criterion tests. The model gives a
best-fit $\chi^2$ value of 210 for 192 degrees of freedom (dof). The
GoF of this is 18\%. The flat cosmological constant model has a best-fit 
$\chi^2$ value of 195 for the same number of dof, giving a GoF of
44\% (Table~\ref{t:aic}). Thus, comparing the GoF for the models may not seem to warrant a strong preference for one model over the other. 
However, since the models
have the same number of fitted parameters, the difference in the BIC
is equal to the difference in the $\chi^2$, $\Delta {\rm BIC} \approx 14$ (see Table~\ref{t:aic} and Fig.~\ref{fig:daic}),
indicating a strong preference for the flat cosmological constant 
model. 

In order to assess the significance of the IC results, we have performed Monte Carlo tests with 1000 simulated data
sets where the underlying cosmology is given by the flat DGP.  We
fitted flat DGP and flat $\Lambda$ models to each simulated data set and then compared the BIC for the two
fits. The $\Delta {\rm BIC}$ obtained\footnote{$\Delta {\rm BIC}={\rm BIC}^{\rm flat \Lambda} - {\rm BIC}^{\rm flat DGP}$.} is roughly
Gaussian with $\Delta {\rm BIC} = -19\pm 9$.  So if the underlying universe genuinely followed the DGP model, we would expect our measured $\Delta {\rm BIC}$ to be negative.  Instead we find a $\Delta {\rm BIC}$ of $+14$.  The highest value we obtain from our 1000 simulated DGP data sets is
$\Delta {\rm BIC} = 7$, still far from the value of 14 we obtain for the real data.
When the underlying cosmology is assumed to be flat $\Lambda$, our simulated data sets give $\Delta {\rm BIC}
= 18\pm 8$, fully compatible with the measured value. From this we conclude that a large difference in the IC of different models indeed points to a very strong statistical preference for the model with the lower value of the IC.

\begin{figure}
   \plotone{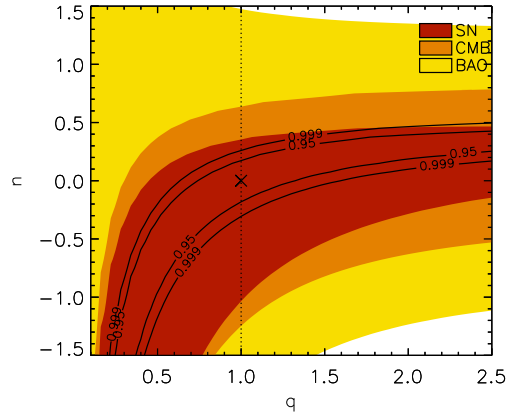}
   \caption{Cardassian expansion (Sect.~\ref{sect:cardassian}).  The dotted line shows the parameters that would agree with the flat, constant-$w$ model (for a wide range of $w$-values).  The cross marks the parameters that match the flat $\Lambda$ model.  This is a model with three free parameters ($\om$ is not shown), and it is not very well constrained by the current data.} \label{cacomb3}   
   \label{fig:cardassian} 
\end{figure}

\begin{figure}
   \plotone{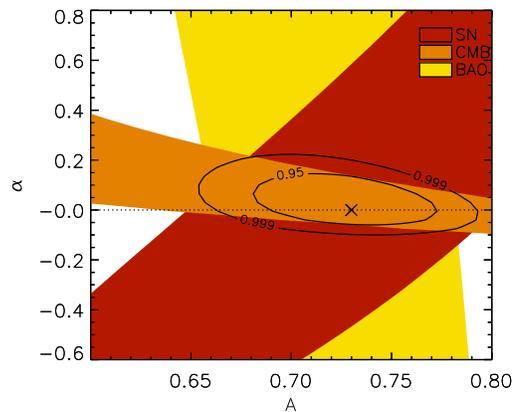}
   \caption{Flat generalized Chaplygin gas (Sect.~\ref{sect:generalchaplygin}).  The cross at $\alpha=0$, $A=0.72$ marks the parameters that match the best-fit flat $\Lambda$ model, while the dotted line shows the parameters that match the $\Lambda$ model (with $\om=1-A$).  Again, despite the flexibility of this model, the best fit is achieved for parameters that are consistent with the flat $\Lambda$ model.} 
   \label{fig:chfcomb} 
\end{figure}
\subsection{Cardassian Expansion}\label{sect:cardassian}\label{sect:ca}

Cardassian models \citep{freese02} involve a modification of the Friedmann equation that allows for  acceleration in a flat, matter-dominated universe.  The reason for the modification could be the self-interaction of dark matter, or the embedding of our observable three-dimensional brane in a higher-dimensional universe.  \citet{wangfreese03} calculated that a Cardassian model could be discriminated from a generic quintessence model or a cosmological constant given expected future data sets, such as the {\em SuperNova Acceleration Probe} \citep[{\em SNAP};][]{snap04}, one possible manifestation of the Joint Dark Energy Mission.  We now employ the interim data sets that have become available to see what current data can determine.

Note that we have assumed that any non-standard modifications to the location of the CMB and BAO peaks are negligible. 

The original power-law Cardassian model has,
\begin{equation} {H^2 \over H_0^2}={\om\over a^3}+{\ok\over a^2}+{(1- \om- \ok)\over  
a^{3n}}\, ,\end{equation} 
where $n$ is a dimensionless parameter related to $w$.  The original Cardassian model is equivalent to the standard dark-energy model (Sect.~\ref{sect:standardde}) 
for $w=n-1$, so there is no need to additionally fit that model.  
However, other incarnations of Cardassian expansion do not match any standard dark-energy model.  One example is ``modified polytropic Cardassian'' expansion, which follows,
\begin{equation} {H^2 \over H_0^2}={\om\over a^3}
\left(1+{(\om^{-q}-1)\over a^{3q(n-1)}}\right)^{1 \over q} \, .\end{equation} 
For $q=1$, this collapses to the flat dark-energy model with  
$w=n-1$.  

Cardassian expansion fits the data well. 
This is due to its close phenomenological similarity with standard dark-energy models.  In particular, we note that the best-fit Cardassian expansion parameters are consistent with (within $1\sigma$ of) those that make the Cardassian expansion collapse to one of the standard dark-energy models (see Fig.~\ref{fig:cardassian}).  
However, it suffers in AIC and BIC tests because of its larger number of parameters (three).


\begin{figure}
   \plotone{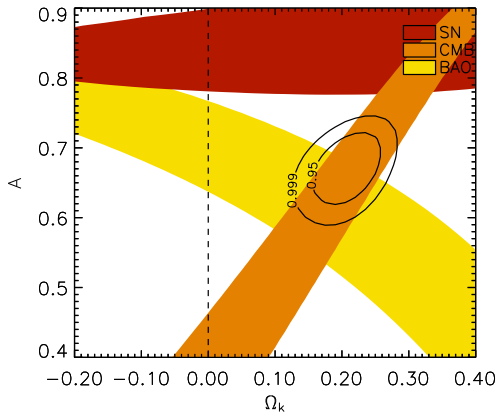}
   \caption{Standard Chaplygin gas (Sect~\ref{sect:stdchaplygin}).  The dashed line shows the flat version of the model.  Clearly this model is a very poor fit to the data.  The subtleties of information criteria are not required to determine that this model is disfavored.} 
   \label{fig:schcomb} 
\end{figure}

\vspace{8mm}
\subsection{Chaplygin gas}
Chaplygin gas models \citep{kamenshchik01} invoke a background fluid with $p\propto \rho^{-\alpha}$.   
They are motivated by brane-world scenarios \citep[][ and references therein]{bento02} and may be able to unify dark matter and dark energy \citep{bilic02}.  We consider both the generalized and standard ($\alpha=1$) Chaplygin gas models, with and without flatness constraints.  

\subsubsection{Generalized Chaplygin Gas}\label{sect:generalchaplygin}\label{sect:gch}
The generalized Chaplygin gas has an equation of state governed by $p=-A/\rho^\alpha$ (with $\rho>0$ and $A$ being a positive constant) and obeys
\begin{equation} {H^2 \over H_0^2}={\ok\over a^2}+(1-\ok )\left( A+{(1-A)\over a^ 
{3(1+\alpha )}}\right)^{1\over 1+\alpha} \, ,\end{equation}
where the standard cosmological constant model is recovered for $\alpha = 0$ and $\om = (1-\ok)(1-A)$.
The reduced distance to the last-scattering surface has been calculated as 
\begin{equation} R=\sqrt{\frac{(1-\Omega_k)(1-A)}{|\ok|}}\,S_k\left[H_0\sqrt{|\ok|}\int_0^{z_{\rm ls}}\frac{dz}{H(z)}\right]. \end{equation}
No modifications for the location of the BAO peak have been made.  
The flat version requires $\ok=0$.  

Out of all the non-standard cosmological models that we consider, Chaplygin gas models fare the best  under the information criteria tests (see Table~\ref{t:aic}), with the flat version slightly preferred (Fig.~\ref{fig:chfcomb}).  This is not unexpected, as the best-fit parameters are again close to flat $\Lambda$ \citep[as also found in, e.g., ][]{bean03}.

\subsubsection{Standard Chaplygin Gas}\label{sect:stdchaplygin}\label{sect:ch}
The standard Chaplygin gas ($\alpha = 1$) has
\begin{equation} {H^2 \over H_0^2}={\ok\over a^2}+(1-\ok )
\sqrt{A+{(1-A)\over a^6}} \, .\end{equation}
Again we also test the flat version, which requires $\ok=0$.
These standard Chaplygin gas models may be the most basic models arising from d-brane theory, but they are not good fits to the data \citep[Fig.~\ref{fig:schcomb}; see also][]{bean03,zhu04}.

\begin{figure}
\plotone{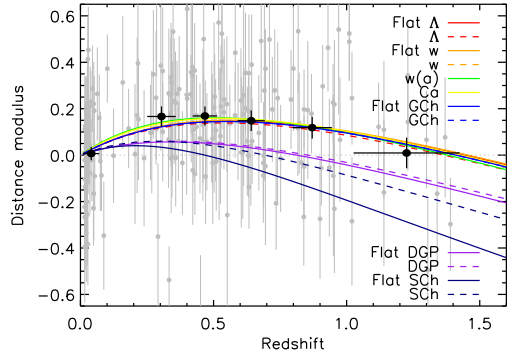}
\caption{Best fit for each model, determined using supernova, BAO and CMB data, plotted against the supernova data.  Distance modulus is shown with respect to the empty model.  Here you can see that the models that provide good fits to the data ({\em lines in the upper legend}) are all essentially identical fits, while the models that could not mimic a cosmological constant fit poorly ({\em lines in the lower legend}).  The grey points in the background are all the raw supernova data used in the fits, while for illustrative purposes only we show the binned data as large black filled circles with two dimensional error bars.  Following \citet{riess07} we use bins of $n\Delta z=6$, where $n$ is the number of points in the bin and $\Delta z$ is the redshift range.  Distance modulus error bars are the quadrature sum of the distance modulus uncertainties in the bin, while redshift error bars show the standard deviation of the redshifts in the bin.}
\label{fig:hd}
\end{figure}
\begin{figure}
\plotone{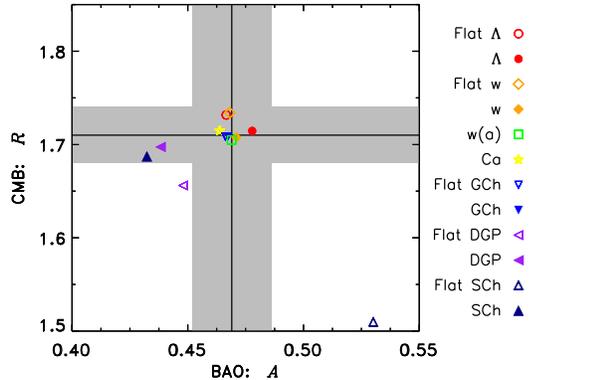}
\caption{Best-fit values of the CMB $R$ parameter and the BAO $A$ parameter for each model.  The solid lines and grey shaded regions show the measurement and $1\sigma$ uncertainty for these parameters as measured using CMB or BAO.}
\label{fig:cmb_bao}
\end{figure}

\begin{deluxetable}{lcccc}
\tablewidth{0pc}
\tablecolumns{5}
\tablecaption{Summary of the Information Criteria Results}
\tablehead{
\colhead{}  &
\colhead{$\chi^2$/ dof } & 
\colhead{GoF (\%)} &
\colhead{$\Delta$AIC} &
\colhead{$\Delta$BIC}
}
\startdata
  Flat cosmo. const. & 194.5 / 192 & 43.7 &   0 &   0 \\
Flat Gen. Chaplygin & 193.9 / 191 & 42.7 &   1 &   5 \\
Cosmological const. & 194.3 / 191 & 42.0 &   2 &   5 \\
 Flat constant $w$ & 194.5 / 191 & 41.7 &   2 &   5 \\
         Flat w(a) & 193.8 / 190 & 41.0 &   3 &  10 \\
      Constant $w$ & 193.9 / 190 & 40.8 &   3 &  10 \\
    Gen. Chaplygin & 193.9 / 190 & 40.7 &   3 &  10 \\
        Cardassian & 194.1 / 190 & 40.4 &   4 &  10 \\
               DGP & 207.4 / 191 & 19.8 &  15 &  18 \\
          Flat DGP & 210.1 / 192 & 17.6 &  16 &  16 \\
         Chaplygin & 220.4 / 191 &  7.1 &  28 &  30 \\
    Flat Chaplygin & 301.0 / 192 &  0.0 &  30 &  30 \\
\enddata
\tablecomments{The flat cosmological constant (flat $\Lambda$) model is preferred by both the AIC and the BIC.  The $\Delta$AIC and $\Delta$BIC values for all other models in the table are then measured with respect to these lowest values.  The goodness of fit (GoF) approximates the probability of finding a worse fit to the data.  The models are given in order of increasing $\Delta$AIC.}
\label{t:aic}
\end{deluxetable}

\section{Discussion and summary}\label{sect:results}\label{sect:discussion}

We have tested a variety of non-standard cosmological models against the latest cosmological data. This includes new data of SNe~Ia from the ESSENCE, SNLS, and Higher-$z$ collaborations. We have also included the reduced distance to the last-scattering surface from the CMB and the constraints from BAO.   
Based on information criteria, both AIC and BIC, the simplest model of a flat universe with a cosmological constant remains the best model to explain the current data.  

Information criteria provide a valuable way to get a relative ranking of the viability of scenarios, using a statistical analysis that gives strong weight to the most simplistic model that fits the observations.
 This does not mean that the simplest model is always correct, rather that more complex and flexible models are not (yet) necessary.  A poor information criteria result will arise when the data are not good enough to adequately constrain the model.  
In order to falsify a model one should look for contradictions in the data -- such as when multiple data sets measure inconsistent values for the parameters of the model.   This occurs, for example, in the standard Chaplygin gas model (Fig.~\ref{fig:schcomb}).

\begin{figure}
\plotone{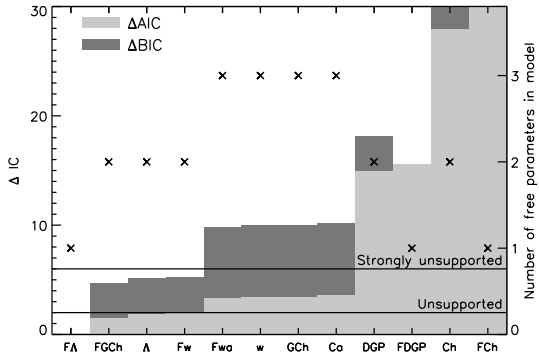}
\caption{Graphical representation of the results in Table~\ref{t:aic}.  $\Delta$AIC is represented by the light grey bars, $\Delta$BIC by the dark grey.  The order of models from left to right is the same as the order in Table~\ref{t:aic}, which is listed in order of increasing $\Delta$AIC.  The crosses mark the number of free parameters in each model ({\em right-hand ordinate}).  The AIC and BIC qualitatively agree on the ranking of models, but the BIC is harsher than the AIC on models with more parameters.  
The primary increase in $\Delta$IC for the first seven models (after flat $\Lambda$) is due to the larger number of parameters, whereas the last four models suffer because they are poor fits to the data.  
A $\Delta$BIC of more than 2 (or 6) is considered positive (or strong) evidence against a model \citep{liddle04}.
We express this as ``unsupported'' (or ``strongly unsupported'') by the data.}
\label{fig:daic}
\end{figure}

We provide a graphical representation of the IC results in Figure~\ref{fig:daic}.  This shows not only the $\Delta$AIC and $\Delta$BIC, but also the number of parameters in each model ({\em crosses and right-hand ordinate}).  Given the current data, the flat cosmological constant model is clearly preferred by these tests.  It almost achieves the best fit of all the models despite its economy of parameters.    

Following it are a series of models that give comparably good fits but have more free parameters.  They are flat general Chaplygin gas, cosmological constant, and flat constant $w$, which all have two free parameters; and general Chaplygin gas, flat $w(a)$, Cardassian expansion, and standard dark energy (constant $w$), which have three free parameters.   We show how their magnitude-redshift evolution compares to the supernova data in Figure~\ref{fig:hd} ({\em lines in the upper legend}) and how well they fit the CMB and BAO data in Figure~\ref{fig:cmb_bao}.  Any of these models could eventually prove to be the best description of our universe, but for the moment the data are not sharp enough to demonstrate the value of the extra complexity. 
Nevertheless, it seems suggestive that these models can all reduce to flat $\Lambda$ 
and their best-fit parameters do so (to within $1\sigma$).  The flat general Chaplygin model, for example, reduces to the flat cosmological constant model 
when $\alpha=0$ and $A=1-\om$.  The actual values of the best fit are
$\alpha=0.03\pm0.10$ and $A=0.73\pm0.04$ (corresponding to $\om=0.27\pm0.04$).  

Clearly new and better data are still needed to discriminate between
these models.  The advanced cosmological probes being planned for the
next decade and beyond \citep{DETFastroph06,peacock06} will improve
considerably on current constraints and will be able to vigorously
test the $\Lambda$ model.
If the results of these future experiments remain consistent with
$\Lambda$, it would raise an interesting question.  Is there a point at
which we should accept $\Lambda$ and abandon our scrutiny of more
complex models?   This is particularly problematic when alternative
models are able to match the $\Lambda$ model to arbitrary precision.

The answer is likely to be two-fold.   Observers will continue to
improve on current techniques and may discover new techniques that
could break the degeneracy for some of these models.  Alternatively,
theoretical considerations, such as the discovery of a quantum theory
of gravity that makes accurate predictions in other realms of
physics, may indicate a preference for a particular model.

The last four models we tested, flat DGP, DGP, standard Chaplygin, and flat standard Chaplygin, are clearly disfavored.  They have fewer parameters than models like flat $w(a)$, but they score poorly because they are unable to provide a good fit to the data.  They do not reduce to flat $\Lambda$ for any values of their parameters.  
  
In summary, given the current quality of the data, 
information criteria indicate that there is no reason to prefer any more-complex model over the concordance cosmology, the flat cosmological constant.  It will be exciting to see whether future data sets change this conclusion.

\acknowledgements

ESSENCE is a project funded by NSF grants AST--0443378 and  
AST--0507475.
The DARK Cosmology Centre is funded by The Danish National Research  
Foundation.
J.S.\ and E.M.\ acknowledge support for this study by VR and from the Anna-Greta 
and Holger Crafoord fund. J.S.\ further acknowledges support from Danmarks  
Nationalbank.  A.C.\ acknowledges the support of CONICYT, Chile, under grants FONDECYT 1051061 and FONDAP Center for Astrophysics 15010003. 
B.P.S. acknowledges support of the Australian Research Council.
 T.M.D.\ acknowledges the support of the Villum Kann Rasmussen Fonden and appreciates the hospitality of the University of Queensland during the writing of this paper.

\bibliographystyle{apj}

\end{document}